\documentclass[twoside,twocolumn]{article}
\usepackage[super,sort&compress,comma]{natbib}
\usepackage[version=3]{mhchem} 
\usepackage{times}
\usepackage{sectsty}
\usepackage{balance}
\usepackage{graphicx} 
\usepackage{lastpage}
\usepackage[format=plain,justification=raggedright,
singlelinecheck=false,font=small,labelfont=bf,
labelsep=space]{caption}
\usepackage{fancyhdr}
\usepackage{color}
\usepackage{bm}
\usepackage[right]{eurosym}
\usepackage{epstopdf}
\usepackage{dcolumn}
\usepackage{amsmath,amssymb} 
\usepackage{upgreek}	

\newcommand{\WZ}{\,\ensuremath{\mathrm{cm}^{-1}}}
\newcommand{\meV}{\,\ensuremath{\mathrm{meV}}}
\newcommand{\eV}{\,\ensuremath{\mathrm{eV}}}
\newcommand{\pbar}{\,\ensuremath{\mathrm{bar}}}
\newcommand{\K}{\,\ensuremath{\mathrm{K}}}
\newcommand{\mum}{\,\ensuremath{ {\upmu \mathrm m}}}
\newcommand{\mm}{\,\ensuremath{\mathrm{mm}}}
\newcommand{\mbar}{\,\ensuremath{\mathrm{mbar}}}
\newcommand{\muJ}{\,\ensuremath{ {\upmu \mathrm J}}}
\newcommand{\ns}{\,\ensuremath{\mathrm{ns}}}
\newcommand{\kHz}{\,\ensuremath{\mathrm{kHz}}}
\newcommand{\Angstrom}{\,\ensuremath{\text{\AA}}}
\newcommand{\eg}{e.\,g. }

\begin{document}


\renewcommand{\thefootnote}{\fnsymbol{footnote}}
\renewcommand\footnoterule{\vspace*{1pt}%
\hrule width 3.4in height 0.4pt \vspace*{5pt}}

\makeatletter
\renewcommand\@biblabel[1]{#1}
\renewcommand\@makefntext[1]%
{\noindent\makebox[0pt][r]{\@thefnmark\,}#1}
\makeatother
\renewcommand{\figurename}{\small{Fig.}~}
\sectionfont{\large}
\subsectionfont{\normalsize}

\renewcommand{\headrulewidth}{1pt}
\renewcommand{\footrulewidth}{1pt}
\setlength{\arrayrulewidth}{1pt}
\setlength{\columnsep}{6.5mm}
\setlength\bibsep{1pt}

\twocolumn[
  \begin{@twocolumnfalse}
\noindent\LARGE{\textbf{Imaging spectroscopy of rubidium atoms attached to helium nanodroplets}}
\vspace{0.6cm} 

\noindent\large{\textbf{L. Fechner\textit{$^{a,b}$}, B. Gr{\"u}ner\textit{$^{a}$}, A. Sieg\textit{$^{a}$}, C. Callegari\textit{$^{c}$}, F. Ancilotto\textit{$^{d}$}, F. Stienkemeier\textit{$^{a}$}, and M. Mudrich\textit{$^{a,e}$}}}\vspace{0.5cm}

\noindent\textit{\small{\textbf{Received Xth XXXXXXXXXX 20XX,
Accepted Xth XXXXXXXXX 20XX\newline
First published on the web Xth XXXXXXXXXX 200X}}}

\noindent \textbf{\small{DOI: 000}}
\vspace{0.6cm}

\noindent \normalsize{Highly excited states of rubidium (Rb) atoms attached to helium (He) nanodroplets are studied by two-photon ionization spectroscopy in combination with electron and ion imaging. Photoion spectra and angular distributions are in good agreement with a pseudodiatomic model for the RbHe$_N$ complex. Repulsive interactions in the excited states entail fast dissociation followed by ionization of free Rb atoms as well as of RbHe and RbHe$_2$ exciplexes. Nonresonant excitation generates Rb atoms in low-lying electronic levels by statistical evaporation-like desorption.}
\vspace{0.5cm}
 \end{@twocolumnfalse}
  ]


\footnotetext{\textit{$^{a}$~Physikalisches Institut, Universit\"at Freiburg, 79104 Freiburg, Germany}}

\footnotetext{\textit{$^{b}$~Present address: Max-Planck-Institut f\"ur Kernphysik, 69117 Heidelberg, Germany}}

\footnotetext{\textit{$^{c}$~Sincrotrone Trieste, 34149 Basovizza, Trieste, Italy}}

\footnotetext{\textit{$^{d}$~Dipartimento di Fisica ``G. Galilei'', Universit\`{a} di Padova, 35131 Padova, Italy and CNR-IOM-Democritos National Simulation Center, 34014 Trieste, Italy}}

\footnotetext{\textit{$^{e}$~E-mail: mudrich@physik.uni-freiburg.de}}

\section{Introduction}
Superfluid helium nanodroplets (He$_N$) doped with alkali (Ak) metal atoms are intriguing quantum systems at the border between gas-phase and condensed matter physics. Due to their extremely weak binding to He droplets Ak atoms reside in dimple-like states at the surface of He nanodroplets.~\citep{Mayol:2005,Dalfovo:1994,Ancilotto:1995,Stienkemeier2:1995} Since the electronic degrees of freedom are essentially centered on the metal atom, electronic excitation induces a strong local perturbation to the equilibrium position of the atoms entailing a complex relaxation dynamics of the AkHe$_N$ complex. This includes excitation of bulk and surface modes of He$_N$ followed by energy dissipation by evaporation of He atoms, spin and electronic relaxation of the dopant Ak atom, the formation of metastable AkHe$_n$, $n=1,2,\dots$ exciplexes, and eventually the desorption of neat Ak atoms or AkHe$_n$ complexes off the He$_N$ surface. The vigorous AkHe$_n$ interaction initiated by laser-excitation of the dopant Ak atom manifests itself in absorption spectra as considerable shifts and broadenings of spectral lines of up to $\sim 1000\WZ$.~\citep{Buenermann:2007,LoginovPhD:2008,Loginov:2011}

The spectra of the lowest electronic excitations are well reproduced by modeling the doped He droplets as pseudo-diatomic molecules where the Ak dopant constitutes one atom and the entire He droplet the other.~\citep{Mayol:2005,Stienkemeier2:1995,Buenermann:2007,Pifrader:2010,Loginov:2011,Callegari:2011} In this picture, which does not account for the internal degrees of freedom of the He droplets, upon electronic excitation the weakly bound ground state of the AkHe$_N$ complex undergoes bound-free transitions to the mostly repulsive short-range part of the pseudo-diatomic potential. Broad absorption bands reflect the widths of the Franck-Condon regions determined by the overlap of the ground state and excited state continuum wave functions. The short-range repulsion readily explains the observed desorption of excited Ak atoms Ak$^*$ off the He surface following excitation. Exceptions are rubidium (Rb) and cesium (Cs) atoms in the lowest excited state ($5\text{P}_{1/2}$ and $6\text{P}_{1/2}$, respectively) that remain attached to He$_N$ due to a shallow outer potential well and a potential barrier at intermediate AkHe$_N$ distance.~\citep{Auboeck:2008,Theisen:2010,Theisen:2011} While Ak$^*$He$_N$ interactions are mostly repulsive, Ak$^*$He pair potentials feature both repulsive and attractive branches. Hence, during the desorption of the Ak$^*$ atom from the He$_N$ surface, metastable Ak$^*$He and Ak$^*$He$_2$ exciplexes can form upon laser excitation. The final population of rovibrational states of the free Rb$^*$He$_N$ molecules is determined by the states of Rb$^*$He$_N$ that are initially populated by the excitation laser as well as by vibrational relaxation taking place in the course of the desorption process due to energy dissipation into the He droplets.~\citep{Reho:2000,Reho2:2000,Mudrich:2009,Gruner:2011,LoginovPhD:2008}

The pseudo-diatomic model (PDM) for electronic transitions to states with low principal quantum numbers is supported by recent time-resolved measurements of dispersed fluorescence emission induced by laser-excitation of Rb attached to He droplets.~\citep{Pifrader:2010} There, the fluorescence decay was found to be consistent with spontaneous emission rates of bare atoms that had quickly desorbed from the droplets such that droplet-induced relaxation needed not be invoked. The PDM was recently impressively confirmed by ion images recorded when exciting sodium (Na) atoms attached to He droplets to either the perpendicular $\Sigma$-state or to the parallel $\Pi$-component with respect to the droplet surface correlating to the Na $4\text{P}$ level.~\citep{LoginovJPCA:2011} The anisotropic ion-momentum distributions were in remarkable agreement with the angular distributions of products of one-photon photodissociation of a diatomic molecule. Photoionization spectra of NaHe$_N$ were interpreted using the PDM for excitations to states with principal quantum numbers up to $n\leq6$. At higher levels of excitation $n>6$, the AkHe$_N$ interaction energies become comparable with the Ak level separations which induces mixing of electron configurations. This state mixing entails electronic relaxation of the excited atoms into lower lying levels. At even higher principal quantum numbers, the doped He droplet evolves into an unusual Rydberg system that features a Rydberg series converging to the vertical ionization potential of the doped He droplet.~\citep{Loginov:2011,LoginovJPCA:2011}

In the present paper we study the excited states $4\text{D}$, $6\text{S}$, $6\text{P}$ of Rb atoms attached to He nanodroplets by means of one-color ns-pulsed resonant two-photon photoionization (R2PI). Time-of-flight mass spectra of the photoions evidence the ejection of bare Rb atoms as well as of RbHe and RbHe$_2$ exciplexes upon electronic excitation. In addition, velocity-map imaging (VMI) is used for visualizing the photoion and photoelectron distributions when resonantly ionizing via the $6\text{P}\Pi$-state. Anisotropic ion images confirm the diatom-like dissociation dynamics of the RbHe$_N$ complex as for the case of NaHe$_N$, whereas narrow lines in the photoelectron spectra clearly reveal that ionization takes place after the Rb atoms have desorbed off the He droplets. Surprisingly, considerable yields of photo ions and electrons are observed between the RbHe$_N$ bands due to the excitation of extended blue wings of lower-lying bands.

\begin{figure}
\begin{center}{
\includegraphics[width=0.48\textwidth]{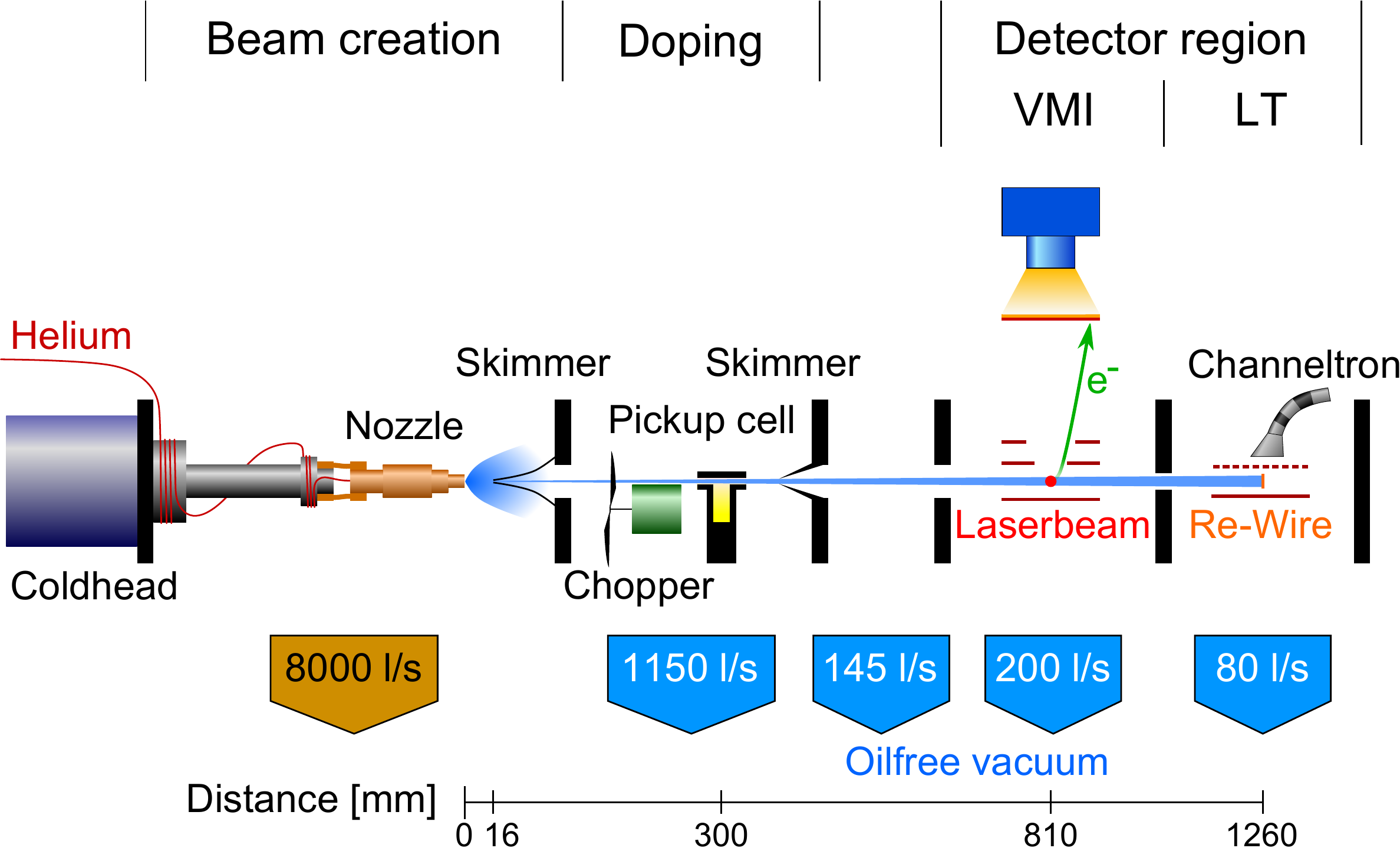}}
\caption{Scheme of the experimental setup used to create a doped droplet beam which is crossed by the laser beam inside the VMI-spectrometer (see text).}
\label{fig:ExpSetup}
\end{center}
\end{figure}

The experimental setup used to produce a beam of Rb-doped helium nanodroplets is schematically represented in Fig.~\ref{fig:ExpSetup} and is very similar to previously used setups~\citep{Mudrich:2004,Claas:2006a,Claas:2007,Mudrich:2009} except for the ionization and detection schemes. Ultrapure He gas is expanded at high pressure ($\sim50\pbar$) through a cold nozzle ($T\approx22\K$, diameter $d=5\mum$) into vacuum. At these expansion conditions the average size of the He droplets amounts to $\left\langle N\right\rangle\approx2500$ He atoms.~\citep{Toennies:2004,Stienkemeier:2006} The droplets enter the adjacent doping chamber through a skimmer ($d=400\mum$) where they pick up single Rb atoms on their way through a stainless steel pickup cell containing Rb vapor. Further downstream, the doped droplet beam passes an empty chamber used as differential pumping section and is then crossed by a pulsed dye laser beam inside the VMI spectrometer. For beam analysis purposes, a Langumir-Taylor (LT) detector is attached to the end of the beam line.~\citep{Stienkemeier:2000}
\section{Experimental}

The VMI spectrometer consists of electrodes arranged in the standard Eppink-Parker configuration~\citep{Eppink:1997}, a grounded flight tube and a position sensitive MCP detector in chevron geometry (Photonis) with an active area that has a diameter of $40\mm$. Single ion and electron hits on the MCP are detected by a phosphor screen and a CCD-camera (Basler / Scout). The recorded images are analyzed by numerical inverse Abel transformation using the BASEX package in order to retrieve the full three-dimensional information about speeds and directions of ions or electrons.~\citep{dribinski} During the experiments, the pressure in this chamber was better than $2\times10^{-8}\mbar$. A pulsed dye-laser (Sirah / Cobra) provides pulses of energy up to $10\muJ$ and of duration of $8\ns$ at a repetition rate of up to $1\kHz$. A Fresnel rhomb is used to rotate the laser polarization parallel or perpendicular to the spectrometer axis before it is focussed onto the droplet beam by a lens with a focal length of $f=200\mm$. For technical reasons, the mass-resolved measurements were performed using a standard TOF mass spectrometer without position sensitivity.

\section{Photoionization spectra}
\label{sec:PISpectra}
\begin{figure}
\begin{center}{
\includegraphics[width=0.48\textwidth]{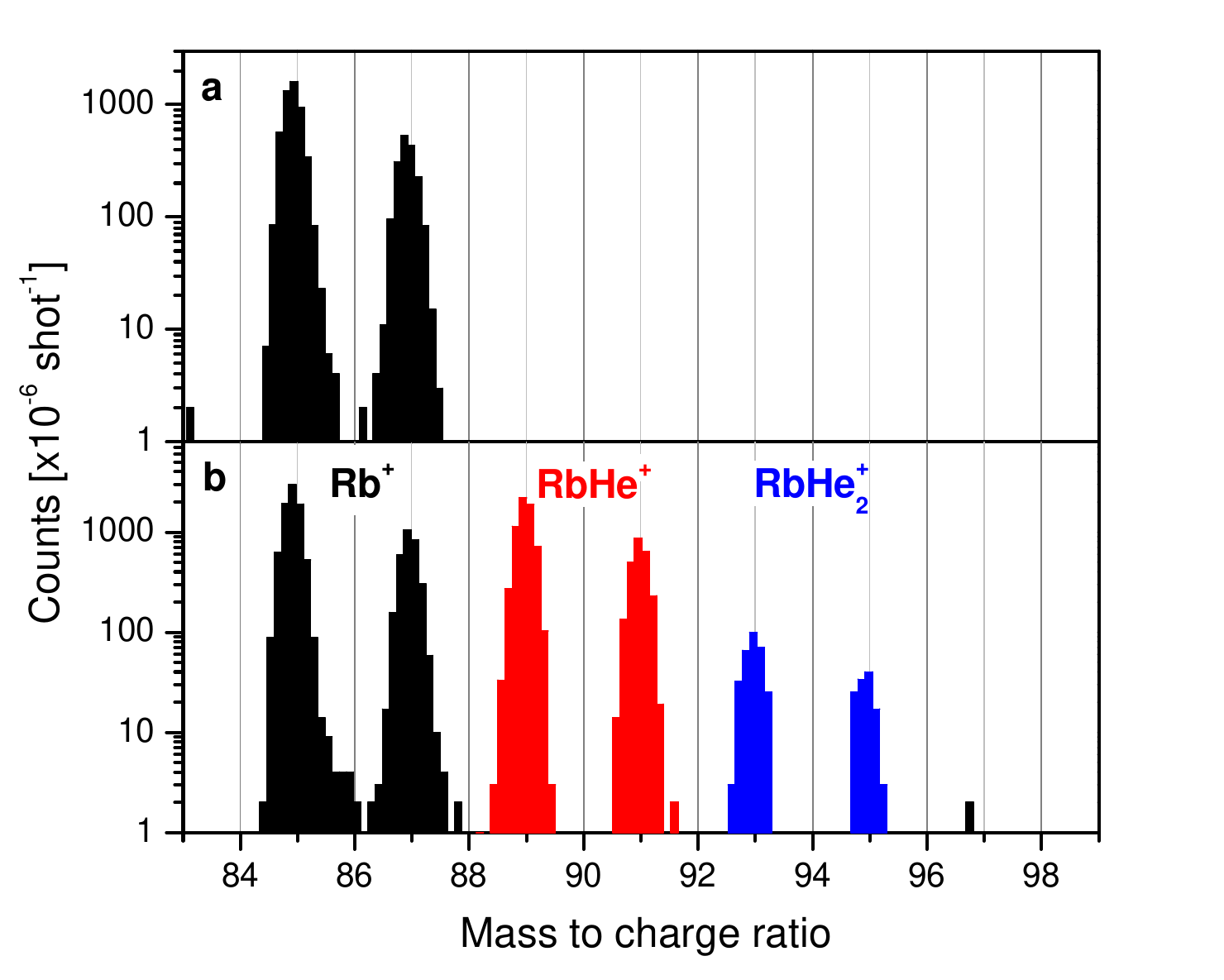}}
\caption{Time-of-flight mass spectra of rubidium (Rb) atoms in the gas-phase photoionized at a wavenumber of $23792.6\WZ$ (a) and of Rb atoms attached to helium nanodroplets photoionized at $24050\WZ$ (b).}
\label{twospectra}
\end{center}
\end{figure}
Gas-phase Rb can easily be ionized using a R2PI process $5\text{S}_{1/2}\rightarrow 6\text{P}_{3/2}\rightarrow \text{Rb}^+$ at $23792.591\WZ$.~\citep{sansonetti}
The corresponding mass spectrum depicted in Fig.~\ref{twospectra}~(a) shows two peaks representing the stable isotopes $^{85}$Rb and $^{87}$Rb with the natural abundance ratio $\sim0.72/0.28$. The same experiment performed with Rb-doped He nanodroplets brings up more interesting mass spectra. Apart from Rb ions, ionized exciplexes RbHe$_n$, $n=1,2$, which are stable molecules only in excited states, appear in the mass spectrum with intensities up to those of neat Rb atoms (Fig.~\ref{twospectra} (b)). Following laser excitation, Rb atoms or exciplexes mostly desorb off the droplets and get ionized by a second photon from the same laser pulse. The timescale for desorption is assumed to fall into the range of a few picoseconds, much shorter than the pulse duration.~\citep{Reho:1997,takayanagi,Droppelmann:2004,Mudrich:2008} Only Rb in the $5\text{P}_{1/2}$-state was found to remain bound to the droplet surface with high probability which gives rise to high yields of Rb$^+$-He$_N$ snowball complexes upon ionization.~\citep{Theisen:2010}

\begin{figure}
\begin{center}{
\includegraphics[width=0.48\textwidth]{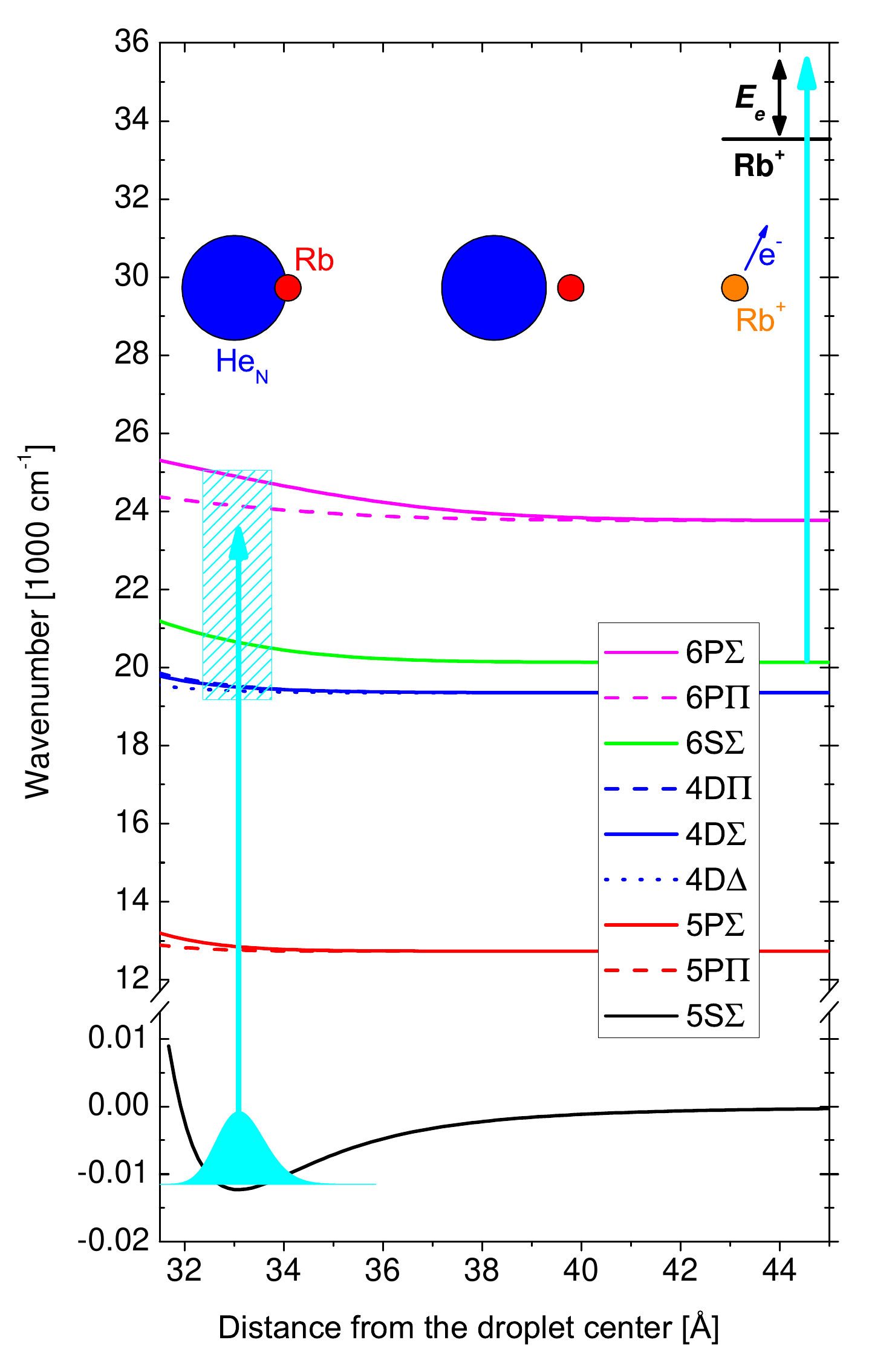}}
\caption{Potential energy curves of the RbHe$_{2000}$ complex from Ref.~\citep{Callegari:2011}.
The Gaussian-shaped curve in the bottom potential represents the vibronic ground state wave function. The arrows indicate the excitation and ionization steps that occur within one laser pulse. The shaded area illustrates the energy range probed in this work.}
\label{fig:potentials}
\end{center}
\end{figure}
The relevant potential energy curves (PEC) of the RbHe$_{2000}$ complex in the PDM are depicted in Fig.~\ref{fig:potentials}.~\citep{Callegari:2011} 
The nearly Gaussian-shaped line in the well of the lowest $5\text{S}\Sigma$-potential represents the ground state nuclear wave function of Rb located at a distance of $6.4\Angstrom$ away from the droplet surface. The left arrow and the shaded area indicate the range of excitation accessed by the first photon in this study. The right arrow symbolizes the ionization step that takes place after dissociation of the RbHe$_N$ complex and $E_e$ stands for the kinetic energy of the ejected photoelectron. Note that the lowest excited states correlating to the two $5\text{P}_{1/2,\,3/2}$ fine-structure levels of Rb have been extensively studied previously.~\citep{Auboeck:2008,Bruehl:2001,Droppelmann:2004,Mudrich:2008,Theisen:2010}

Most importantly, all of the represented PECs are well separated from each other such that no significant state-mixing is expected in the course of the nuclear dynamics initiated by laser excitation, \eg by level crossings. Furthermore, all PEC except for the ground state are purely repulsive. Therefore we may expect the Rb atoms to efficiently desorb off the He droplets upon excitation. Due to the splitting of the asymptotic P-states into $\Sigma$ and $\Pi$ components associated with perpendicular and parallel orientations of the P-orbital with respect to the He droplet surface, respectively, and to the splitting of the D-states into $\Sigma$, $\Pi$, and $\Delta$ projections, we expect to observe a total of $5$ bands (4D$\Sigma,\,\Pi$, 6S$\Sigma$, 6P$\Sigma,\,\Pi$ in in the range $19000\dots25000\,$cm$^{-1}$ according to the selection rule for one-photon E1-transitions $\Delta\Lambda = 0,\pm1$. Since spin-orbit coupling is relatively weak ($\lesssim78\,$cm$^{-1}$) for the levels studied in this work we completely neglect it. Note that although the atomic transitions $5\text{S}\rightarrow 4\text{D},\,6\text{S}$ are forbidden, the corresponding transitions of droplet-bound Rb, $5\text{S}\Sigma\rightarrow 4\text{D}\Sigma ,\,4\text{D}\Pi ,\,6\text{S}\Sigma$ become allowed due to breaking of the atomic symmetry by the He surface.

\begin{figure}
\begin{center}{
\includegraphics[width=0.48\textwidth]{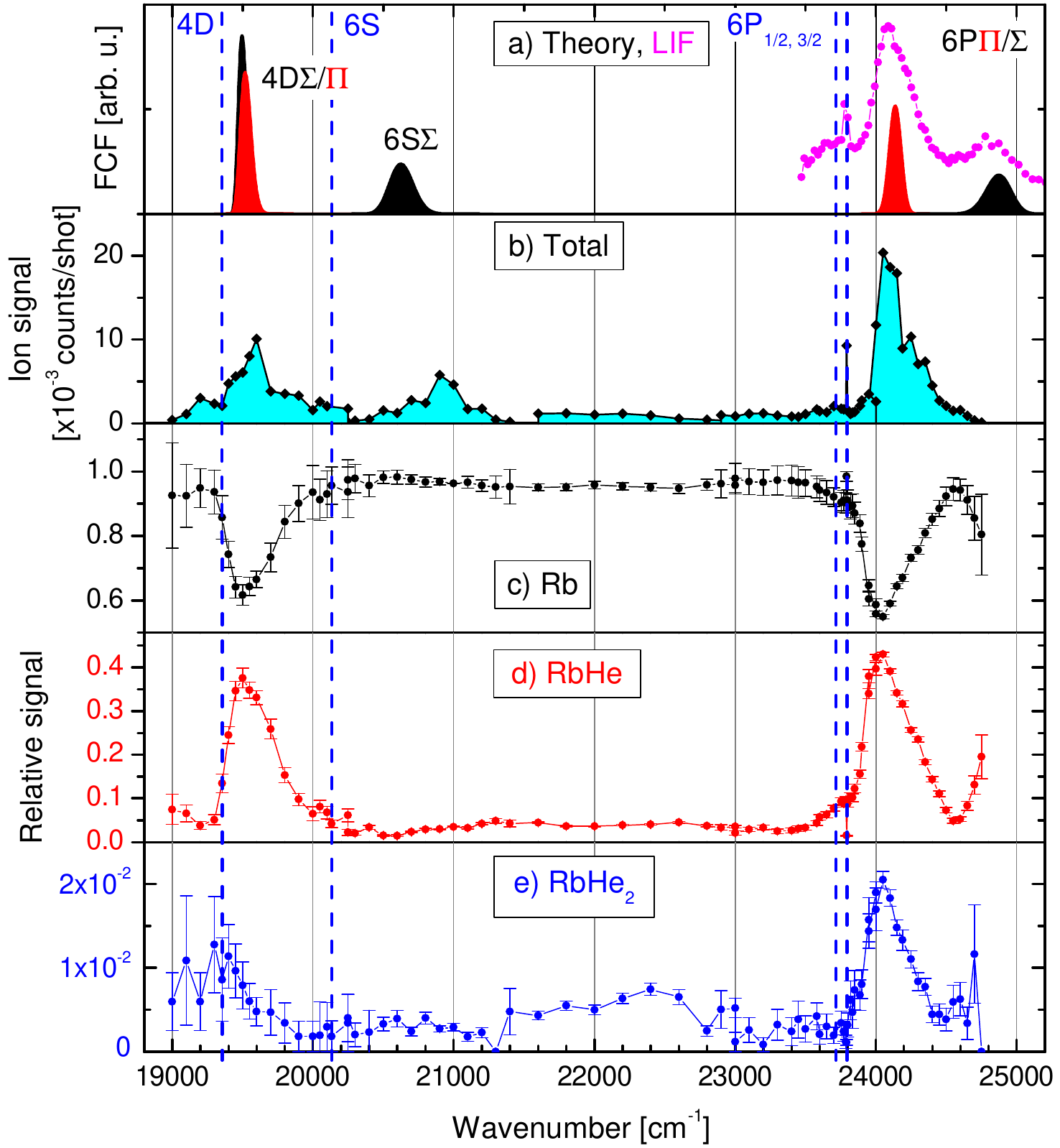}}
\caption{Simulated (a) and measured (b) photo ionization spectrum of rubidium (Rb) atoms attached to helium nanodroplets. The symbols in (a) represent laser-induced fluorescence (LIF) data from Ref.~\citep{Pifrader:2010}.
The relative signal contributions of neat Rb, and of exciplexes RbHe and RbHe$_2$ are shown in (c)-(e). The dashed lines depict the atomic transition energies.}
\label{fig:verlauf}
\end{center}
\end{figure}
A synthetic PI spectrum is generated by computing Franck-Condon factors of the transitions from the vibronic ground state using LeRoy's program BCONT 2.2~\citep{bcont} and is displayed in Fig.~\ref{fig:verlauf}~(a). In addition the symbols represent measurements of laser-induced fluorescence (LIF) extracted from Ref.~\citep{Pifrader:2010}.
The measured integral PI spectrum, shown in Fig.~\ref{fig:verlauf}~(b), is recorded using several different laser dyes that produce variable laser pulse energies. For this reason, and due to the fact that the excited states have different ionization cross sections and feature different saturation behavior, a quantitative comparison of the intensities of spectral features is not possible. While the positions of the $4\text{D}\Sigma,\,\Pi$ and $6\text{P}\Pi$ bands match the simulated ones and the LIF spectrum reasonably well, the $6\text{S}\Sigma$ bands is significantly blue-shifted. A similar discrepancy was also found for the corresponding $4\text{S}\Sigma$ band of Na attached to He droplets, whereas a different model calculation resulted in a slight red shift of the measured peak.~\citep{LoginovJPCA:2011} Therefore we attribute this shift to deficiencies of the present PDM potentials which may be related to slightly differing droplet sizes used in the experiment ($\left\langle N\right\rangle\approx2500$) and for the PDM ($N=2000$). The large scattering of the experimental data points in Fig.~\ref{fig:verlauf}~(b) is mainly due to instabilities of the dye laser intensity which partly factor out when calculating signal ratios (Fig.~\ref{fig:verlauf}~(c)-(e)).

Similarly to the case of Na on He droplets~\citep{LoginovJPCA:2011}, the measured features are considerably broadened with respect to the calculated ones. Several processes may contribute to additional line broadening: Spin-orbit splitting of the bands correlating to $4\text{D}$ and $6\text{P}$ atomic states that are not included in the model calculation, inhomogeneous broadening due to averaging over the broad distribution of He droplet sizes, and He density fluctuations.~\citep{Buenermann:2009,Hernando:2010} Besides, the calculated line widths may be underestimated due to the limited accuracy of the used model potentials. An important source of line broadening is saturation due to the high laser intensity which easily occurs in such experiments using pulsed lasers for spectroscopy. Since in the employed one-color R2PI scheme the first excitation step has by far higher cross sections than the second ionization transition, the former transition is saturated. The dependence of the yield of photoions as a function of laser pulse energy significantly deviates from quadratic scaling in particular at high pulse energies. This is in spite of using comparably low pulse energies $\lesssim10\muJ$. In addition to saturation broadening of the RbHe$_N$ bands associated with the PDM, additional phonon wings can be present that extend each band far to the blue.~\citep{Hartmann:1996} Such additional spectral features result from the excitation of He droplet degrees of freedom that are not considered in the PDM. These phonon wings feature very different saturation behavior than the PDM bands. By this we explain the observation of photo ions and electrons far away from the RbHe$_N$ bands, \eg between the $6\text{S}\Sigma$ and $6\text{P}\Pi$ bands at $\sim23000\WZ$. The corresponding photo ion and electron images are discussed in Sec.~\ref{sec:ions} and \ref{sec:electrons}, respectively.

From the time-of-flight mass spectra recorded at each wavenumber we infer the relative contribution of neat Rb atoms and of exciplexes RbHe$_n$ to the integral photoion signal, shown in Fig.~\ref{fig:verlauf}~(c)-(e). Substantial exciplex formation occurs only in the $4\text{D}$ and the $6\text{P}$ bands of the RbHe$_N$ complex, whereas nearly no exciplexes are observed at the $6\text{S}\Sigma$-resonance and in between the bands. RbHe$_2$ exciplexes are significantly formed only in the $6\text{P}\Pi$ band and at the very red edge of the $4\text{D}\Sigma,\,\Pi$ bands. While these findings are similar to those obtained with Na atoms attached to He droplets~\citep{LoginovPhD:2008,LoginovJPCA:2011}, they differ in certain details. In both cases excitation into the lowest excited $\text{S}\Sigma$ bands ($4\text{S}\Sigma$ and $6\text{S}\Sigma$ for Na and Rb, respectively) generates only free atoms and no exciplexes. In the case of NaHe$_N$, however, NaHe as well as NaHe$_2$ exciplexes were measured with large abundance in the lowest D band ($3\text{D}\Sigma,\,\Pi$) and only very small amounts of NaHe and no NaHe$_2$ were seen in the second P band ($4\text{P}\Pi$). In contrast to that, we measure most efficient RbHe and significant RbHe$_2$ exciplex formation in the Rb $6\text{P}\Pi$ band, whereas in the Rb $4\text{D}$ band only small amounts of RbHe$_2$ occur.

\begin{figure}
\begin{center}{
\includegraphics[width=0.48\textwidth]{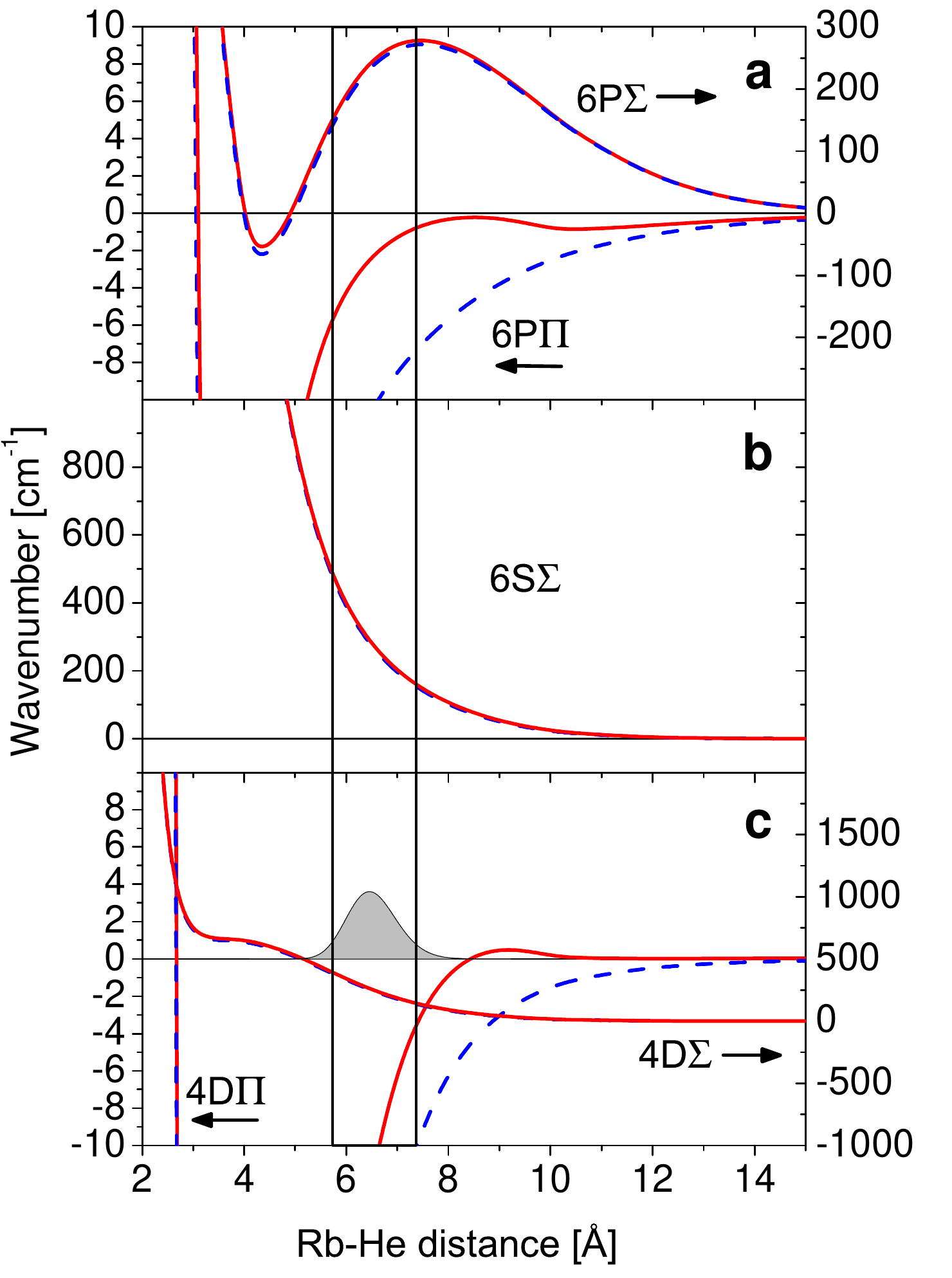}}
\caption{RbHe pair potentials correlating to the $6\text{P}$, $6\text{S}$, and $4\text{D}$ states of Rb (dashed lines). Solid lines represent modified potentials including the He extraction energy out of the He surface for the case of RbHe exciplex formation. The Gaussian-like curve shows the vibronic ground state wave function of the RbHe$_{2000}$ complex.}
\label{fig:Reho}
\end{center}
\end{figure}
A semi-quantitative understanding of the efficiency of RbHe exciplex formation can be obtained from modified RbHe pair potentials taking into account the extraction energy of one He atom out of the He droplet surface to attach to the excited Rb atom. Using the model of Reho \textit{et al.}~\citep{Reho2:2000} and the RbHe pair potentials of Pascale~\citep{Pascale:1983} we obtain the PEC displayed in Fig.~\ref{fig:Reho} (solid lines). The dashed lines represent the unmodified potentials for comparison. In the cases of purely attractive PECs ($6\text{P}\Pi$, $4\text{D}\Pi$), the inclusion of the He abstraction energy leads to a potential barrier at $R\approx8\Angstrom$ that principally could prevent the exciplex formation if excitation occurred at large distances $R\gtrsim8\Angstrom$. In this case the formation rates can be estimated using a tunneling model through that barrier, as done for NaHe and KHe exciplexes.~\citep{Reho2:2000} In our case of RbHe, however, vertical excitation out of the equilibrium state of Rb at the surface of a He droplet of size 2000 -- the RbHe$_{2000}$ vibronic ground state wave function -- leads to direct formation of RbHe in bound vibrational states when exciting into $6\text{P}\Sigma,\,\Pi$, and $4\text{D}\Pi,\,\Delta$-states. The PECs of $6\text{S}\Sigma$ and $4\text{D}\Sigma$ are purely repulsive such that no exciplex formation is expected.

These conclusions are indeed qualitatively confirmed by the experimental findings: efficient RbHe formation in the $6\text{P}$ and $4\text{D}$ bands, no RbHe in the $6\text{S}\Sigma$ band. Since the $4\text{D}\Sigma,\,\Pi$ bands partly overlap, a clear distinction as to the relative RbHe formation efficiencies in $\Sigma$ and $\Pi$-states is not possible. The fact that the RbHe yield is largest in the $6\text{P}\Pi$ band may be related to the shape of the modified $6\text{P}\Pi$ potential that features only a shallow barrier which stays below the dissociation limit. In contrast, the height of the barrier of the modified $4\text{D}\Pi$-potential surpasses the $4\text{D}$ dissociation threshold which principally allows vibrationally highly excited RbHe to dissociate by tunneling outwards through the barrier.

It should be noted, however, that this static one-dimensional model is oversimplified in several respects. The more complex dimple-shaped He environment allows for many-particle interactions that open other competing dynamical channels such as the formation of larger exciplexes, \eg RbHe$_2$, the excitation of compression and surface modes of the He droplet that are likely to transiently change the local He environment of the excited Rb atom, as well as the desorption of atoms and exciplexes off the droplet surface. In particular the time scale on which desorption proceeds in relation to vibrational relaxation times determines the final distribution of population in the RbHe vibrational levels. This population distribution in turn has an impact on the detection efficiency due to varying photionization cross sections of different vibrational levels.

\section{Photoion images}
\label{sec:ions}
\begin{figure}
\begin{center}{
\includegraphics[width=0.48\textwidth]{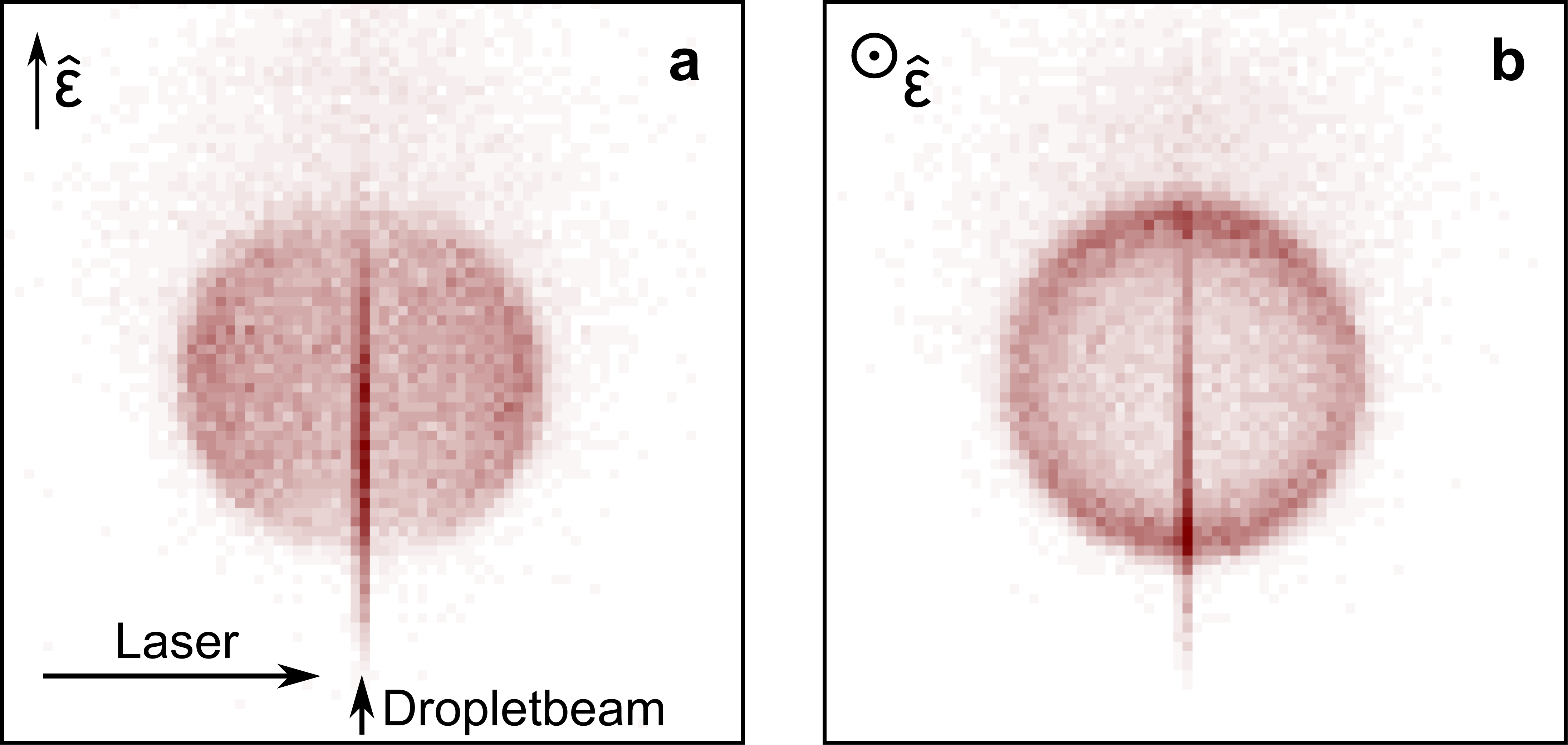}}
\caption{Velocity map images of all photoions generated by photoionization of Rb-doped He nanodroplets at the $6\text{P}\Pi$ band ($23986\WZ$). The laser polarization points perpendicular to the He droplet beam axis and perpendicular (a) and parallel (b) to the symmetry axis of the VMI spectrometer.}
\label{fig:ions}
\end{center}
\end{figure}
The PDM has proven to nicely interpret the excitation spectra of alkali atoms attached to He droplets in the lowest excited states. Furthermore, the fast desorption of most excited states off the He droplets is consistent with this model that mostly predicts repulsive alkali atom-helium PEC. The validity of the PDM is even more directly confirmed by the momentum distribution of photoions Rb, RbHe and RbHe$_2$ recorded by means of VMI. The raw images of all photoions detected without mass selectivity at the laser wavenumber $23986\WZ$, which corresponds to excitation near the $6\text{P}\Pi$ resonance, are shown in Fig.~\ref{fig:ions} for the laser polarization $\hat{\varepsilon}$ pointing perpendicular (a) and parallel (b) to the VMI spectrometer axis of symmetry. The narrow central streak stems from Rb atoms propagating in an effusive beam along the He droplet beam axis which are non-resonantly ionized. The strongly elongated shape of this distribution reflects the large ratio of velocities in longitudinal versus transverse directions with respect to the atomic beam axis due to the high degree of beam collimation. The second more extended circular component in Fig.~\ref{fig:ions}~(a) and (b) represents ionized Rb atoms and exciplexes that have desorbed off the He droplets upon excitation prior to ionization. In contrast to the effusive beam signal this distribution is not notably broadened along the beam direction owing to the sharp longitudinal velocity distribution of the He droplet beam. Note that Rb and RbHe ions produce identical ion images irrespective of their different masses provided they have equal kinetic energies.

\begin{figure}
\begin{center}{
\includegraphics[width=0.48\textwidth]{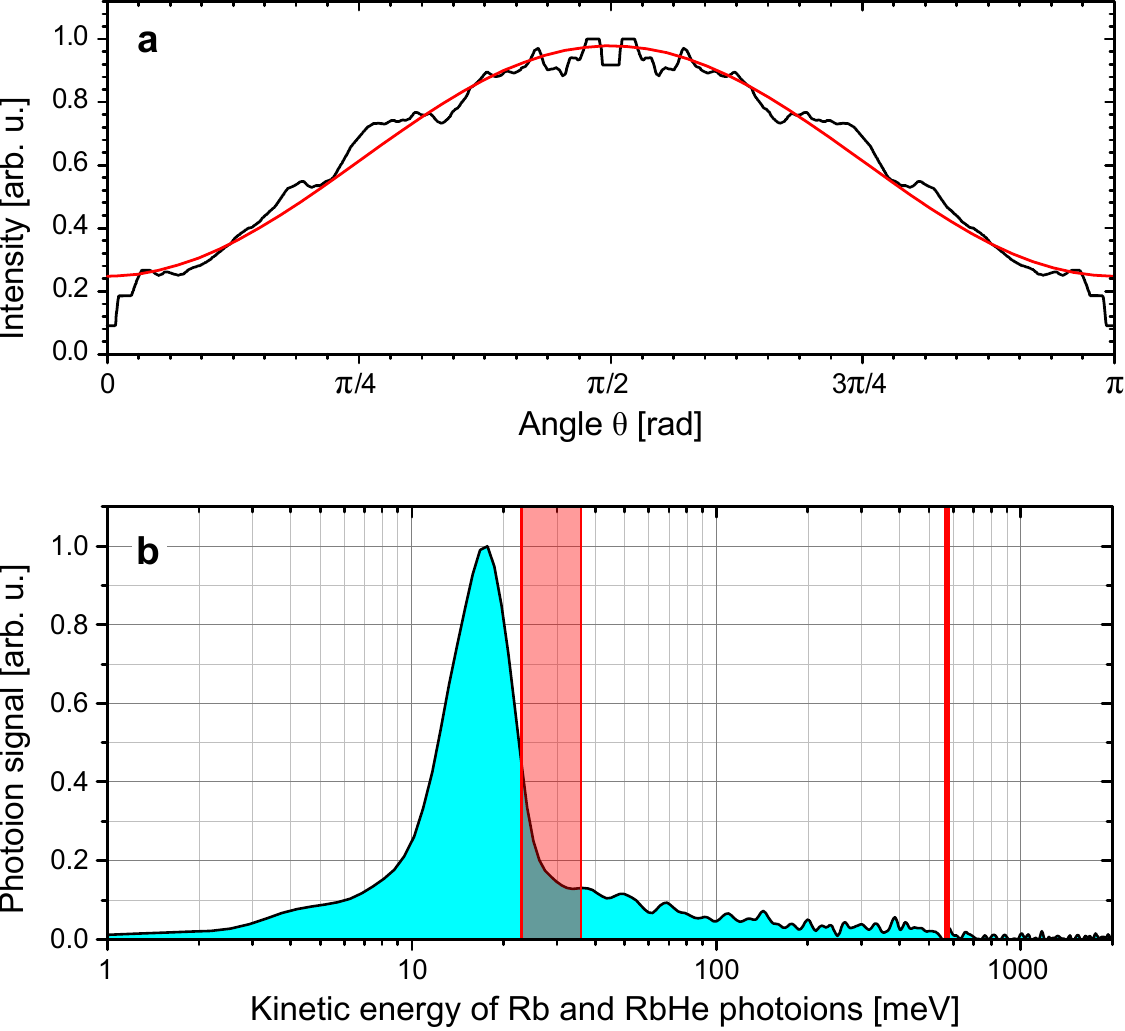}}
\caption{Angular intensity distribution (a) and kinetic energy spectrum of photoions (b) retrieved from the inverse Abel transformed ion image shown in Fig.~\ref{fig:ions}.}
\label{fig:2spektren_ionen}
\end{center}
\end{figure}
For symmetry reasons, only the distribution of Fig.~\ref{fig:ions} (a) can be inverse Abel transformed to retrieve the angular intensity distribution $I(\theta)$ and ion kinetic energies $E_{Rb}$. From the transformed image we obtain $I(\theta)$ by integrating over the radius and the total kinetic energy spectrum of atomic and exciplex ions by integrating over the angle $\theta$. We find that $I(\theta)$ nicely matches the characteristic intensity distribution for a perpendicular $\Sigma\rightarrow\Pi$ dissociative transition in a diatomic molecule (Fig.~\ref{fig:2spektren_ionen} (a)). By fitting the well-known expression~\citep{Zare:1972} $I(\theta)\propto 1+\beta P_2(\cos\theta)$
we obtain a value for the anisotropy parameter $\beta=-0.6631(11)$ which is close to the ideal value $\beta=-1$. This result nicely confirms the validity of the PDM and the assignment of the spectral band to the perpendicular transition $5\text{S}\Sigma\rightarrow 6\text{P}\Pi$.

The ion spectrum shown in Fig.~\ref{fig:2spektren_ionen} (b) features a pronounced peak around $E_{Rb}=17.3\meV$ that is shifted to lower energies with respect to the excess energy 
that the ions would acquire upon PDM-like dissociation at the given laser wavenumber (left vertical bar in Fig.~\ref{fig:2spektren_ionen} (b)). The He droplets are assumed as hard spheres with the mass of $N$ He atoms. The effective mass of the collision partner of the Rb atoms and RbHe molecules in the dissociation reaction can be inferred by assuming a two-body hard-sphere half collision and considering energy and momentum conservation.~\citep{LoginovPhD:2008} This yields the number of He atoms out of the He dimple that effectively interact with Rb (RbHe), $N_{\mathrm{eff}}=\frac{m_{Rb}}{m_{He}}\frac{E_{Rb}}{E_D-E_{Rb}}$. Here, $m_{Rb}$ denotes the mass of Rb (RbHe) and the total energy release $E_D=24.0$\,meV ($36.1$\,meV) is given by the difference between photon energy and internal energy of the products, Rb atoms in the $6\text{P}$-state (RbHe in $6\text{P}\Pi$ vibrational levels). Since the degree of vibrational relaxation upon formation of RbHe is not known we estimate a binding energy of $12.8$\,meV based on the potential curves by Pascale.~\citep{Pascale:1983} Presumably all vibrational states of RbHe are populated due to partial relaxation, as observed for NaHe and KHe.~\citep{Reho:2000} Thus, the energy release is augmented by binding energies ranging from 0 to $12.8$\,meV, which may contribute to the broadening and shifting of the peak in Fig.~\ref{fig:2spektren_ionen} (b) to energies $\gtrsim24.0$\,meV.
Surprisingly, we find $N_{\mathrm{eff}}\approx56$ ($N_{\mathrm{eff}}\approx21$) for Rb (RbHe), which is much larger than the values $1.4\lesssim N_{\mathrm{eff}}\lesssim5.7$ found for Na in excited states $3\text{P},\,4\text{S},\,3\text{P},\,4\text{P}$.~\citep{LoginovPhD:2008}
This may be related to the slightly stronger binding of Rb atoms to the He droplets~\citep{Ancilotto:1995,Mayol:2005} as well as to the larger charge radius of excited Rb atomic orbitals as compared to those of Na. A quantitative understanding, however, requires precise modeling of the dynamics of the whole RbHe$_N$ system, as it is currently being performed for Na atoms attached to He droplets.~\citep{Barranco:privateCom}

In addition to the prominent peak the ion spectrum in Fig.~\ref{fig:2spektren_ionen} (b) features an extended pedestal reaching out to kinetic energies $E_{Rb}\lesssim0.5\eV$. Although no definite conclusion can be drawn as to the origin of this signal component at this stage, we suggest that it reflects desorption dynamics involving more complex processes that are beyond the PDM, \eg statistical evaporation-like desorption induced by massive excitation of He droplet modes. This assumption is backed by the photoelectron spectra discussed in the following section which reveal that lower-lying electronic states correlating to $4\text{D}$ and $5\text{P}_{3/2}$ atomic levels can be excited far off the PDM bands. Consequently, the maximum kinetic energy available to the Rb atoms (RbHe exciplexes) that are ejected from the He droplets mostly in the 4D ($\Sigma,\,\Pi$) states amounts to the difference between photon energy and the 4D term energy which is $0.574\eV$ ($0.586\eV$) (right vertical bar in Fig.~\ref{fig:2spektren_ionen} (b)). This value nicely matches the signal onset in the ion spectrum.

As the laser wavenumber is tuned below the $6\text{P}\Pi$ band the anisotropic ion images gradually become nearly isotropic and the signal intensity distributes smoothly over roughly the same range of energies $E_{Rb}=0\dots 0.5\eV$. However no distinct ring-shaped structure is apparent. We speculate that the structures in both angular and energy distributions are lost due to statistical isotropic desorption becoming the dominant mechanism. Moreover, the fact that several asymptotic Rb states ($6$P, $4$D, $5$P) contribute to the signal additionally tends to wash out the angular structure of the ion images.

\section{Photoelectron spectra}
\label{sec:electrons}
Further insight into the dynamics of the excitation process of Rb attached to He droplets is obtained from imaging the photoelectron distributions. Experimentally, photoelectron images are obtained very easily just by reversing the polarity of the voltages applied to repeller and extractor electrodes. Let us start by exemplifying the photoelectron images obtained by resonantly ionizing free Rb atoms from the effusive beam. Since the lifetime of the excited $6\text{P}_{3/2}$-state is known to be much longer ($\sim109\ns$)~\citep{heavens} than the laser pulse length, the population of this state is not significantly affected by spontaneous decay. Thus, all photoelectrons are emitted with a kinetic energy $E_e=2h\nu - E_{IP}\approx1.72\eV$. Here, $h\nu=2.91\dots 2.97\eV$ denotes the photon energy and $E_{IP}=4.1771\eV$ is the ionization potential of Rb.

\begin{figure}
\begin{center}{
\includegraphics[width=0.48\textwidth]{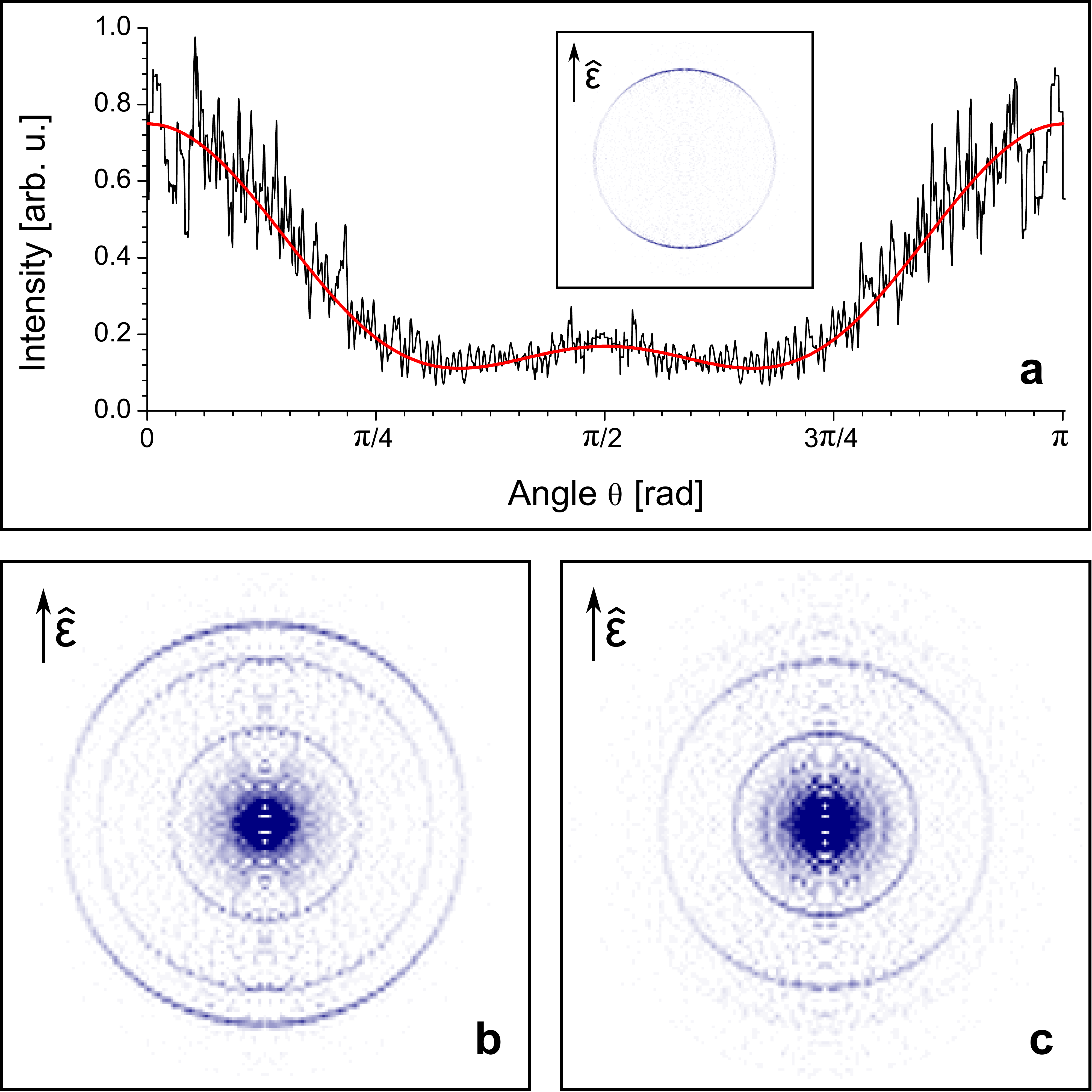}}
\caption{Inverse Abel transformed photoelectron images recorded by ionizing bare Rb atoms (a) and Rb-doped He nanodroplets (b). The main graph in (a) represents the angular intensity distribution including a fit curve from which anisotropy parameters are extracted. The two images in (b) and (c) are recorded at laser wavenumbers $23985\WZ$ and $23753\WZ$, respectively.}
\label{fig:RbAngle}
\end{center}
\end{figure}
An inverse Abel transformed photoelectron image of free Rb atoms ionized by R2PI at $23792.6\WZ$ is shown in the inset of Fig.~\ref{fig:RbAngle}~(a). Since all created electrons have the same kinetic energy, the signal is concentrated in one single ring with a radius defined by $E_e$. Furthermore, a preferred direction of emission along the polarization of the laser is clearly visible. A closer inspection reveals additional slightly enhanced emission probability perpendicular to the polarization axis. The corresponding angular distribution obtained by radial integration is depicted in the main plot of Fig.~\ref{fig:RbAngle}~(a). As shown in Ref.~\citep{reid},
the photoelectron distribution resulting from the absorption of two equally linearly polarized photons can be described by an expression containing two anisotropy parameters $\beta_2$ and $\beta_4$ and Legendre polynomials of second and fourth order, $I(\theta)\propto1+\beta_2P_2(\cos(\theta))+\beta_4P_4(\cos(\theta))$.
From a fit of this expression to the experimental data (smooth line in Fig.~\ref{fig:RbAngle}~(a)) we obtain $\beta_2=1.280(9)$ and $\beta_4=0.987(10)$. These values are in good agreement with earlier results from simple field-free time-of-flight-measurements ($\beta_2=1.30(8)$ and $\beta_4=1.13(8))$.~\citep{cuellar}

The situation changes when using Rb-doped He nanodroplets instead of gas-phase Rb. Since the $5\text{S}\Sigma\rightarrow 6\text{P}\Pi$ band of RbHe$_N$ is blue-shifted with respect to the atomic $5\text{S}\rightarrow 6\text{P}$-transition by about $250\WZ$ the photoelectron imaging measurements we have performed in the range $23750\dots 23950\WZ$ are actually located in the red edge of this band. Surprisingly, under these conditions new rings appear in the photoelectron images, as shown in Fig.~\ref{fig:RbAngle}~(b), (c). In addition to the outer ring, which coincides with the one seen in the gas-phase image associated with electrons ejected out of the $6\text{P}$-state, two new sharp rings and a bright central spot appear.

\begin{figure}
\begin{center}{
\includegraphics[width=0.48\textwidth]{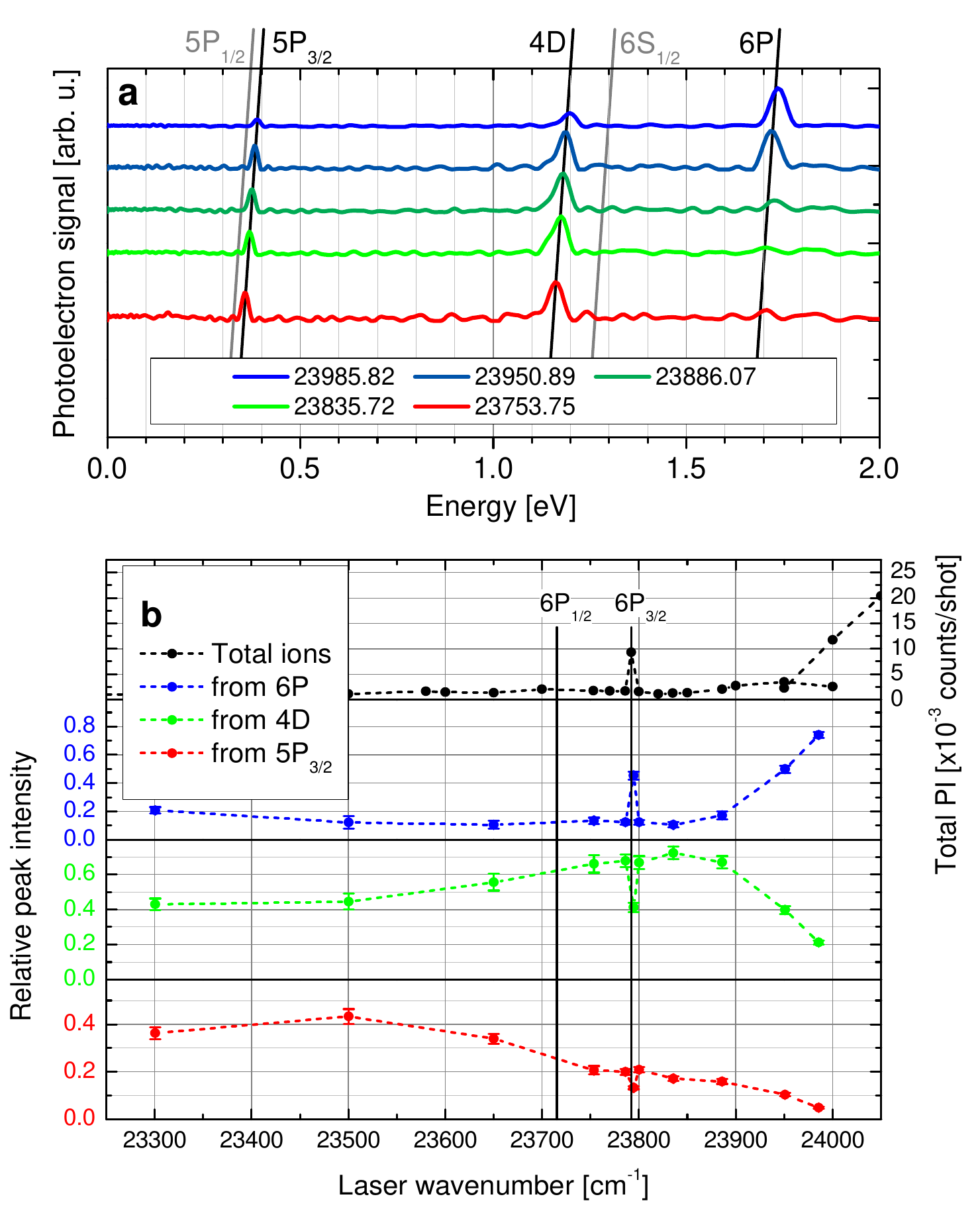}}
\caption{(a) Photoelectron spectra inferred from photoelectron images recorded at various laser wavenumbers. The three peaks can be assigned to electrons ejected out of states $5\text{P}_{3/2}$, $4\text{D}$ and $6\text{P}$ of bare Rb atoms. (b) Relative yields of electrons out of the three states as a function of laser wavenumber. For comparison the upper graph shows the ion spectrum (cf. Fig.~\ref{fig:verlauf} (b)).}
\label{fig:electronsspectra}
\end{center}
\end{figure}

The photoelectron spectra extracted from the images recorded at various laser wavenumbers are depicted in Fig.~\ref{fig:electronsspectra}~(a) together with the total ion-yield from Fig.~\ref{fig:verlauf}~(b) for comparison. Note that these spectra are normalized so that only the relative peak intensities within each spectrum can be analyzed. The energy axis is calibrated to the atomic line in the spectrum of gas-phase Rb in the effusive beam. The widths of the peaks are instrument limited. Clearly the three visible peaks can be assigned to electrons ejected out of lower-lying levels of free Rb, namely $5\text{P}_{3/2}$ and $4\text{D}$. The fine-structure splitting of the $4\text{D}$-state is not resolved.

The relative amplitudes of the three photoelectron peaks as a function of the laser wavenumber are shown in Fig.~\ref{fig:electronsspectra}~(b). Thus, in the range between the atomic $5\text{S}\rightarrow 6\text{P}$-transition at $23792.6\WZ$ and the peak of the corresponding RbHe$_{N}$ band at about $24100\WZ$ photoelectrons from the $6\text{P}$-state have the largest share. At larger red-detunings the $6\text{P}$-peak drops down to a constant signal level that we ascribe to electrons produced by non-resonantly ionizing Rb atoms from the effusive beams that are also visible in the ion images (see Fig.~\ref{fig:ions}). The two peaks corresponding to electrons out of $5\text{P}_{3/2}$ and $4\text{D}$ remain with roughly equal relative intensities. We find the following anisotropy parameters which hardly change in the studied range of laser wavenumbers, $6\text{P}$: $\beta_2=1.18(18)$ and $\beta_4=0.05(7)$; $4\text{D}$: $\beta_2=0.97(11)$ and $\beta_4=0.1(2)$; $5\text{P}_{3/2}$: $\beta_2=0.58(7)$ and $\beta_4=0.0(1)$. Surprisingly, $\beta_4$ is equal to zero within the experimental uncertainty for all Rb atoms that have detached from the He droplets, irrespective of their final state. This means that the initial orbital alignment induced in the excitation step is quenched by the desorption process. In recent measurements of the photoelectron emission following EUV-IR pump-probe photoionization of pure He droplets, vanishing $\beta_4$-parameters were observed for spectral features associated with the excitations inside the diffuse surface region that are prone to strong interactions with the surrounding He and subsequently undergo intraband relaxation.~\citep{Kornilov:2011} In contrast, He Rydberg atoms expelled from the outermost surface region retained their original orbital alignment. Apparently, excited alkali atoms interact more strongly with the He environment despite their surface location.

Two possible scenarios are conceivable that interpret the nonzero signal at excitation wavenumbers below $23700\,$cm$^{-1}$ and the occurrence of desorbed free Rb atoms in low-lying levels: either excitation of a red wing of the 5S$\Sigma \rightarrow$6P$\Pi$ band or excitation of a blue wing of the 5S$\Sigma\rightarrow$4D$\Sigma$, and 5S$\Sigma\rightarrow$5P$\Sigma$ bands.

The first case has been observed in the spectrum of the 2S$\rightarrow$2P transition in He nanodroplets doped with lithium (Li) atoms and has been related to a lateral displacement of the Li atom along within the dimple where it sits.~\cite{Buenermann:2007}
We dismiss this scenario as unlikely to explain our data, primarily because of the considerable range that this hypothetical red wing must span (compare the shoulder of the above-mentioned Li transition, which is short and has a very abrupt onset). It would also disagree with measurements similar to ours in Na-doped He droplets.~\cite{LoginovPhD:2008,LoginovJPCA:2011} Let us however illustrate the implications of its occurrence, which may be relevant in the immediate vicinity of the 6P$\Pi$ band: conservation of energy means that the Rb atom must either remain on the droplet (in which case it will hardly be detected by R2PI and photoelectron spectroscopy, but very easily by laser-induced fluorescence spectroscopy), evaporate as an exciplex, or evaporate by relaxing to a lower electronic state while still on the droplet.
For the sake of reference we note that the expansion coefficients of the 6P$\Pi$-state in terms of atomic states, for $N = 2000$ at the equilibrium distance shown in Fig.~\ref{fig:potentials}, are $88.2\%\, (6\text{P})$, $10.4\%\, (5\text{D})$, $1.1\%\, (4\text{F})$.

On the other hand, the lower RbHe$_N$-states $4\text{D}\Sigma,\,\Pi$ and $5\text{P}\Sigma,\,\Pi$ can be excited directly as a consequence of phonon wings that extend these bands far into the blue. Upon desorption, free atoms in the states $5\text{P}_{3/2}$ and $4\text{D}$ emerge following dissociation of the RbHe$_N$-complex. The fact that such wide phonon wings are efficiently excited may be related to the comparatively high intensities of the laser pulses which are needed for the ionization by a second photon from the same laser pulse to be efficient. The occurrence of phonon wings on the blue side of the absorption bands of doped He droplets which feature quite different saturation behavior compared to the band maxima is well-known.~\citep{Hartmann:1996} This interpretation is further supported by time-resolved measurements of LIF emitted after exciting into the $6\text{P}\Pi$ band.~\citep{Pifrader:2010} The authors concluded that the population of final atomic states was consistent with spontaneous emission rates of the bare atoms such that droplet-induced relaxation needed not be invoked. Furthermore, the ion images we have recorded at laser wavenumbers $\lesssim23800\WZ$ showed broad and isotropic ion energy distributions that speak for the excitation of a superposition of different states and for massive excitation of He droplet modes that induce statistical desorption dynamics rather than PDM-like dissociation.

The missing photoelectron signals from the Rb states $5\text{P}_{1/2}$ and $6\text{S}$ (Fig.~\ref{fig:electronsspectra}~(a)) are currently not fully understood. The lower photoionization cross section of the $6\text{S}$-state by about a factor $10$ as compared to the ones of states $5\text{P}$ and $4\text{D}$ does not fully account for the experimental result.~\citep{Aymar:1984}

The pronounced central feature in the photoelectron images (Fig.~\ref{fig:RbAngle}) stems from electrons with nearly zero electron kinetic energy (ZEKE). Close inspection of this region of the images reveals that the entire signal actually is concentrated in only a few pixels such that only an upper bound for the electron energy $\lesssim10\meV$ can be specified. The fact that the electron energy is very low and independent of the photon energy suggests that the electrons arise from an indirect ionization mechanism involving significant electron-He interactions. The intensity of the ZEKE electron signal is nearly constant at a level of about 2\,\% of the integral electron yield. Therefore these electrons cannot be directly correlated with Rb or RbHe$_n$ photoions, which strongly depend on laser wavenumber. Besides, the ZEKE electron yield shows a pronounced droplet size dependence, in contrast to the electron signals related to the desorbed species. We measure an increase by about a factor of $4$ for the ZEKE electron peak intensity when varying the average size $\left\langle N\right\rangle$ of He droplets from $\left\langle N\right\rangle=300$ to $\left\langle N\right\rangle=3800$. Thus the creation of ZEKE electrons relies neither on RbHe$_N$ resonances nor on the desorption of excited species off the droplets, but seems to be an independent indirect ionization mechanism of the entire RbHe$_N$ complex.

ZEKE electrons have been observed before in experiments using EUV ionization of pure and doped He droplets.~\citep{Peterka:2003,Peterka:2007,Wang:2008} Interpretations range from electrons being transiently trapped in bubble states inside the He which collapse to release slow electrons to unusual Rydberg states that form preferentially at the droplet surface and subsequently decay by autoionization. The fact that definitely no He excitation is involved in our experiments clearly shows that a quite general mechanism is at the origin of the ZEKE electrons. In recent time-resolved EUV photoionization experiment of pure He droplets the occurrence of ZEKE electrons is related to He droplet excitations located in the diffuse surface region of the droplets.~\citep{Kornilov:2011} The fact that the Rb atoms studied in this work are located at the He droplet surface implies that surface-bound electron states most efficiently produce ZEKE electrons. Possibly the ZEKE electrons we measure are correlated with the formation of large cationic snowball complexes~\citep{Theisen:2010} which, however, are too heavy to be detected by our present ion detection scheme.

\section{Conclusion}
In conclusion, we have studied the dynamics of He nanodroplets doped with Rb atoms excited to high electronic states by nanosecond laser pulses in combination with mass-spectrometry and ion and electron imaging detection. Resonant two-photon ionization spectra as well as velocity-mapped ion and electron distributions are in good agreement with the pseudo-diatomic model where the Rb dopant constitutes one atom and the entire He droplet the other. According to this simple picture, the RbHe$_N$ complex is electronically excited by absorption of a first photon, followed by dissociation and subsequent ionization of the free Rb atom by a second photon from the same laser pulse.

Deviations from this model are found when analyzing the recoil kinetic energy of Rb atoms following desorption off the He droplets upon excitation to the $6\text{P}\Pi$ band, from which an effective mass of the He droplet of about $56$ He atoms is inferred. When exciting to bands correlating to the atomic $4\text{D}$ and $6\text{P}$-states, efficient formation of RbHe$_{1,2}$ exciplex molecules is observed in accord with expectations based on RbHe pair potentials.

Furthermore, when exciting off resonance between the $6\text{S}\Sigma$ and $6\text{P}\Pi$ bands we measure significant yields of electrons originating from the lower states $5\text{P}$ and $4\text{D}$. Presumably phonon wings that extend the correlating bands far into the blue are excited as a result of comparatively high laser intensities in the first excitation step of the employed pulsed laser ionization scheme. The quenching of orbital alignment as well as significant production of electrons with nearly zero kinetic energy indicate substantial coupling of the Rb atom to the He droplet prior to desorption. The corresponding ion kinetic energy spectra and angular distributions imply that statistical evaporation-like ejection of Rb atoms takes place in this regime rather than pseudodiatomic dissociation.

These measurements will be extended to higher lying states that are subjected to even stronger configuration interaction mixing induced by the He environment. Resulting non-adiabatic couplings may induce more complex dynamics such as transient solvation of the dopant in the He droplet that may compete with dissociative channels. The intricate dynamics of such conceptually simple benchmark systems will further be investigated in real-time experiments using femtosecond pump-probe spectroscopy.

\section*{Acknowledgments}
We thank M. Drabbels and B. von Issendorff for valuable discussions and advice. Support by the Deutsche Forschungsgemeinschaft (DFG) is gratefully acknowledged.


\providecommand*{\mcitethebibliography}{\thebibliography}
\csname @ifundefined\endcsname{endmcitethebibliography}
{\let\endmcitethebibliography\endthebibliography}{}

\end{document}